\documentclass[%
 reprint,
 amsmath,amssymb,
 aps,
]{revtex4-2}

\usepackage{graphicx}
\usepackage{dcolumn}
\usepackage{bm}

\begin{document}

\preprint{APS/123-QED}

\title{Symmetry emergence in Self-organized criticality}

\author{Ernesto Lupercio}
 \affiliation{CINVESTAV, IMI-BAS}

\author{Mikhail Shkolnikov}
\affiliation{IMI-BAS}

\date{\today}

\begin{abstract}
We describe a mechanism of affine symmetry emergence in the maximal density regime of the prototypical model of self-organized criticality when the inverse square of the mesh of the underlying lattice is much larger than the number of random perturbation points distributed according to a prescribed probability measure supported in the interior of the ambient convex domain. Moreover, an appropriate scaling limit of the toppling function (aka odometer), which counts the number of operations per site, is a solution to a non-linear partial differential equation well known in the context of optimal transport and differential geometry, making it possible to accurately estimate the deviation of the density from its maximal value in any macroscopic window. The mechanism for the affine symmetry emergence is due to the novel empirical fact, supported in addition by inductive arguments that have recently being upgraded to a rigorous proof, that the scaling limit of the toppling function is the unique concave solution of the Monge-Amp\`ere equation with Dirichlet boundary condition on the convex domain with the potential given by the probability measure used above as the infinite-perturbation profile.

\end{abstract}

\maketitle



\medskip
\noindent\textit{Physics Subject Headings.}
Self-organized criticality; nonequilibrium statistical mechanics;
lattice models in statistical physics; critical phenomena;
scaling methods; self-organized systems.

\medskip

	\begin{center}{\footnotesize{\emph{Dedicated to the memory of a prominent thinker, {\bf Ivan Todorov}, who perceived all forms of creative expression as kindred.}}}\end{center}


 ``In the 1960s, which I remember with fondness, there was talk of physicists and lyricists. I think there has always been a kinship at some deep level. Both mathematics and music, and physics and fine arts, and even religion.''

Ivan Todorov, a quote from a 2023 interview \cite{Alexandrova_2025}.

\section{\label{sec:level1}Introduction}

The term ``Self-organized criticality'' refers to a class of complex systems which behave like those at a second-order phase transition but without being finely tuned. It originates from the pioneering work \cite{bak1987self} of Bak, Tang and Wiesenfeld, who proposed the prototype for such a system and studied it numerically. In a generality suitable for our purposes, the adaptation of the setup of their experiment is the following. Consider $\Omega\subset\mathbb{R}^2$ -- a compact convex domain with non-empty interior, and $h>0$ -- the mesh of the square lattice $h\mathbb{Z}^2.$ Take $\Omega_h,$ the collection of sites in the system, to be the intersection $\Omega$ with $h\mathbb{Z}^2.$ A state $\phi$ of the model is an integer-valued function on $\Omega_h,$ it is called stable if $\phi(v)<4$ for each site $v.$

There is a stabilization map $\phi\mapsto\phi^\circ$ sending any state $\phi$ to the unique stable state $\phi^\circ.$ One way to define this map is via the least action principle \cite{fey2010growth}, i.e. among all non-negative integer-valued functions $H$ on $\Omega_h$ such that $\phi+\tilde\Delta H<4,$ where $\tilde\Delta$ is the discrete Laplace operator, there is only one function $H_\phi$ minimizing the action functional $\sum_v H(v).$ We call this function $H_\phi$ the toppling function of $\phi,$ and $\phi^\circ:=\phi+\Delta H_\phi.$ The word ``toppling'' corresponds to the elementary operation $\psi\mapsto\psi+\Delta\tilde\delta_v$ -- toppling at the site $v$ -- where $\tilde\delta_v$ is the function equal to $1$ at $v$ and $0$ otherwise. This operation reduces the value of ``grains'' at $v$ by $4,$ redistributing it among the four neighbors (if $v$ happens to be at the boundary, i.e. some neighbor is missing, the corresponding grain leaves the system). Thus, one may compute the toppling function algorithmically by a procedure known as ``relaxation'': starting with an unstable state $\phi$, perform topplings at sites with values greater than $3,$ in an arbitrary order, until arriving at $\phi^\circ$ -- the toppling function corresponds to the number of topplings per site.

\begin{figure}
    \centering
    \includegraphics[width=1\linewidth]{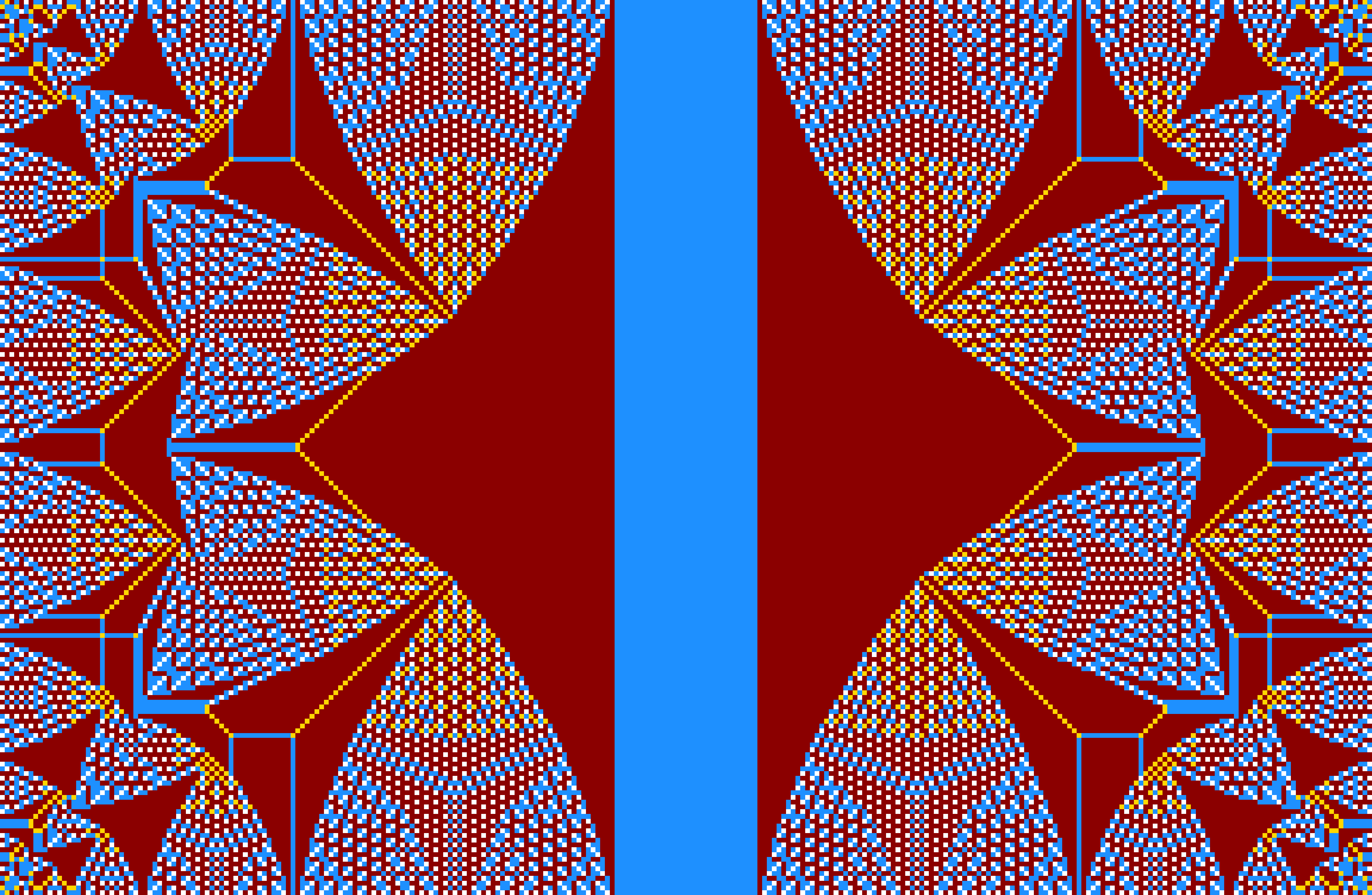}
    \caption{Restoration of the first image of the neutral element of the sandpile group on 288x188 lattice rectangle produced by Michael Creutz \cite{creutz1991abelian}, in the original color scheme: white, gold, dodgerblue and darkred mean 0,1,2 and 3, respectively. Reproduced by Higinio Serrano.}
    \label{fig:Creutz}
\end{figure}

Starting with an arbitrary state $\phi_1,$ Bak, Tang and Wiesenfeld consider a Markov chain $\phi_{k+1}=(\phi_k+\tilde\delta_p)^\circ,$ where each time $p$ is chosen uniformly at random, i.e. we drop grains at random sites at each step, which may produce virtually no effect or result in an avalanche of topplings. Gathering statistics for the sizes of avalanches, they observe power-laws. The states on $\Omega_h$ split into two categories, independent of the particular run of the experiment: transient, which appear at most once; and recurrent, which keep being visited throughout the process with an equal probability. The latter phenomenon is explained by Dhar \cite{dhar1990self} through the notion that soon evolved to be known as the ``sandpile group'' consisting of recurrent states subject to the operation $(\psi_1,\psi_2)\mapsto(\psi_1+\psi_2)^\circ.$ This group is finite Abelian, with the order expressed in terms of spanning forests on $\Omega_h$ with every tree being adjacent to the boundary at a single site, which follows by Kirchhoff matrix-tree theorem on the absolute value of the discrete Laplacian. The neutral element of the sandpile group has a stunning fractal structure, and was first brought to the attention of the public in \cite{creutz1991abelian} by Creutz in the case of $\Omega$ being a rectangle, see Fig. \ref{fig:Creutz}. What one immediately sees is that the neutral element, apart from minor linear defects, is scale invariant.

Indeed, the mentioned power-law is a common signature of scale transcendence manifesting itself which is often possible to express mathematically in the form of a scaling limit theorem. One such theorem \cite{kalinin2026tropical,kalinin2016tropical,kalinin2020pattern} proven by Nikita Kalinin and the second author operates in the maximal density regime, i.e. the initial state for us is $\langle 3\rangle$ taking value $3$ identically on $\Omega_h.$ Varying $h,$ we perturb this state by adding a single grain at sites $p_h$ of $\Omega_h$ near $p$ belonging to a fixed finite subset $P$ of the interior of the $\Omega,$ i.e. look at the state $\phi_{h,P}=\langle 3\rangle+\sum_{p\in P}\tilde\delta_{p_h}$ and its stabilization $\phi_{P,h}^\circ.$ What was discovered is that, as $h\rightarrow 0$, the locus where $\phi_{h,P}^\circ<3$ converges in the Hausdorf sense to a graph $C_{P,\Omega}$ with straight edges passing through all $p$ in $P.$ In fact, the slopes of this edges are all rational, and the so-called balancing condition -- a number-theoretic-flavoured property akin to conservation of momentum or charge -- holds at every vertex. Such graphs are often seen as limits of holomorphic curves and are known under the name ``tropical curves'' in geometry and under the name pq-webs in string theory -- the one appearing in the scaling limit is extremal in the sense of minimizing the symplectic area in the given incidence class, therefore serving as a solution to an analogue of the Steiner problem. 

To be historically accurate, Sadhu and Dhar have foreseen the future relevance of tropical geometry in the study of SOC in \cite{sadhu2012pattern}, and our direct inspiration was the work \cite{caracciolo2010conservation} of Caraciollo, Paoletti and Sportiello, who put forward the notion of ``strings'' in sandpiles, thin, $(p,q)$-periodic patterns with local density deviation expressed as $p^2+q^2$ -- the value exactly corresponding to the area of the holomorphic lift as it was established in \cite{kalinin2020sandpile,kalinin2018introduction}.  As a conceptual twist of \cite{kalinin2018self}, the notion of Tropical Sandpile model is introduced, and its self-organized criticality is demonstrated, thus the prototypical (discrete) SOC converges to the new (continuous) tropical SOC. What was not emphasized before is the emergence of the larger local symmetry, upgrading the eight-element dihedral group $D_4,$ acting through the diagonal and anti-diagonal invertible matrices with non-zero entries equal to $\pm 1$, to the still discrete, but now infinite integer matrix group $\operatorname{GL}_2(\mathbb{Z}).$ Formally, the covariance is expressed as $$C_{A(P),A(\Omega)}=A(C_{P,\Omega})$$ where $A\in \operatorname{GL}_2(\mathbb{Z})$ and $C_{P,\Omega}$ is the tropical curve arising as the scaling limit of perturbations at $P$ on $\Omega.$

In this note, we report on the new twist of the above story. Namely, one can take the further scaling limit of the Tropical Sandpile, arriving to the notion of an ``infinitely perturbed state'' which posses an even larger, now continuous, group of local symmetries -- the full special linear group $\operatorname{SL}_2(\mathbb{R}),$ thus, incorporating the obvious translational symmetry, rendering the model equi-affine invariant. Changing the perspective, one may say that going down from Affine, i.e. macro scale, to Tropical, i.e. meso scale, to Abelian, i.e. micro scale, sandpiles one observes a chain of consecutive symmetry breakings. In addition to the usual bottom-up causation of the scaling limits, one has rudiments of top-down causation, where, despite a theoretical inability to predict values at individual microscopic sites, collective observables, such as a density, an average of microscopic values, in a macroscopic window for a probabilistic perturbation process can be effectively estimated by taking the inverse of the Monge-Amp\`ere operator of the density of perturbation profile and integrating its Laplacian over the window.

\section{From micro to meso scale}
The definition of the Tropical Sandpile model on $\Omega$ is the following. A state of the model is a pair $(F,P),$ where $P$ is a finite subset in the interior of $\Omega$ and $F$ is a tropical series, i.e. an infimum (infinite tropical sum) of affine linear functions with integer gradients (tropical monomials), with the tropical domain of convergence $\Omega,$ i.e. assuming finite value at each point of $\Omega,$ with an additional condition that $F$ vanishes at the boundary, thus taking non-negative values in the interior of $\Omega$ due to concavity. It is a general fact established in \cite{kalinin2018introduction} that such a tropical series $F$ restricted to any compact region in the interior of the domain of its converges is represented by a minimum of a finite number of linear functions, and therefore it makes sense to speak about its corner locus $V(F)$ consisting of all such points where $F$ is not locally linear. 

A state $(F,P)$ is called stable if $P$ is a subset of $V(F).$ For an arbitrary state, there is a stabilization function producing a stable state $(G_P F,P),$ where the point-part of the tuple remains unchanged, and the series-part is characterized by a version of the least action principle. Namely, among all tropical series $E$ with domain of convergence $\Omega$ such that $E\geq F$ and $P$ belongs to $V(E),$ the series $G_P F$ is the unique minimizer of the tropical action functional $\int_\Omega E$ taken with respect to the standard Lebesgue measure. There is a numerical way to find $G_P F$ using iterations of idempotent and mutually non-commuting single-grain operators $G_q=G_{\{q\}}$ for $q$ in $P.$ 
 
This single grain operator for $q$ corresponds to the scaling limit of the microscale operator of adding a grain at a site, stabilizing and removing the grain, i.e.  $$\phi\mapsto\tilde G_v=(\phi+\tilde\delta_v)^\circ-\tilde\delta_v,$$ where $v=q_h$ is a site of $\Omega_h$ near $q$ in $\Omega.$ In the terminology popularized by Vafa and his collaborators, $G_q$ corresponds to what they call ``breathing mode'', which changes the geometry of the tropical curve by shrinking a single face in its complement to which $q$ belongs until the curve passes through $q.$ Crucially, the discrete operators $\tilde G_v$ are naturally decomposed as a power of a wave operator given by $W_v\phi=(\phi+\Delta\tilde\delta_v)^\circ$, i.e. $\tilde G_v \phi=W_v^M\phi$ for some non-negative integer $M$ depending on $\phi.$ Notably, $W_{v_1}\phi=W_{v_2}$ for $v_1$ and $v_2$ belonging to the same cluster of $3$'s on $\phi.$   

Thus, to effectively solve the sandpile model in the maximal density regime, one needs to understand the action of individual waves. The class of states that emerge from the interaction of waves with the boundary are sandpile strings, discussed in the introduction. The edges of which strings are built behave like solitons under the wave action, i.e. move towards the source of the wave linearly with the speed inverse proportional to their density deviation, which is shown in \cite{kalinin2020sandpile}. To establish another version of the Abelian-to-Tropical scaling limit, one may therefore synchronize the microscopic relaxation (among all possible paths, we choose the one that is decomposed into sequences of waves) with the mesoscopic one performed by alternating the continuous operators $G_q$ for all $q$ in $P.$ As a result, for the initial unstable state $(0_\Omega,P),$ where $0_\Omega$ denotes the identically zero tropical series on $\Omega,$ the series-part $G_P 0_\Omega$ of its stabilization is seen as the scaling limit of the $h$-rescaled toppling function for the $P$-perturbed state $\langle 3 \rangle+\sum\tilde\delta_{p_h}$ on $\Omega_h,$ see \cite{kalinin2026tropical} for the proof.

The variational coordinate-free definition of $G_P$ implies the full tropical covariance of the model with local symmetry part being $\operatorname{GL}_2(\mathbb{Z}),$ i.e. $$G_{A(P)} (F\circ A^{-1})=(G_P F)\circ A^{-1},$$ where $A$ is in $\operatorname{GL}_2(\mathbb{Z})$ and $F$ is a tropical series on $\Omega.$ There is a third formulation of micro-to-meso scaling limit, which is capable of restoring the density of the stabilized configuration, but the price to pay is the loss of tropical covariance. Namely, we may look at the quantitative deviation $\langle 3\rangle-(\langle 3 \rangle+\sum\tilde\tilde\delta_{p_h})^\circ$ multiplied by $h^{-1}$ -- it converges to the distribution supported on $C_{P,\Omega}=V(G_P 0_\Omega)$ -- this distribution asymptotically corresponds to the push-forward of the symplectic area of the holomorphic lift of the tropical curve to the complex algebraic torus endowed with the invariant symplectic structure \cite{kalinin2018introduction}.

\section{From meso to macro scale}

What we are doing now conceptually is very simple, namely, we consider the same kind of Markov chain as in the original Bak-Tang-Wisenfeld process, however, not in micro, but in the meso scale, i.e. in tropical sandpile, with an additional twist of taking each perturbation point distributed along a fixed probability distribution with density $\rho$ supported in the interior of $\Omega.$ The main difference is that in the classical setup, the number of recurrent states was finite, thus we are bound to going in cycles, and presently there are no recurrent states at all, so the state steadily grows in complexity, with the total action diverging as $N^{\frac{1}{2}}$ where $N$ is the number of perturbation points -- this asymptotic was originally observed numerically in \cite{kalinin2023some}, and below we will sketch an argument of why in dimension $d$ it is $N^{\frac{1}{d}}.$

As it is often the case, it is instructive to go to the lowest non-trivial dimension first and consider $d=1,$ where the model can be solved explicitly. It is easy to see that for generic $p_1,\dots,p_N,$ negative of the second derivative of $G_{p_1,\dots,p_N}0_{[0,1]}$ is the sum of Dirac delta functions $\delta_{p_1}+\dots+\delta_{p_N}+\delta_q,$ where the additional $q$ in the interval $[0,1]$ is the unique point satisfying the condition of $p_1+\dots p_N+q$ being an integer, see \cite{shkolnikov2023relaxation}. Therefore, assuming that $p_k$ are independent and distributed according to $\rho,$ the function $N^{-1}G_{p_1,\dots,p_N}0_{[0,1]}$ converges to the unique solution of the equation $f''=-\rho$ with boundary conditions $f(0)=f(1)=0.$

What we realized recently in relation to averaging over the space of tropical structures \cite{kalinin2026limits} is that the suitable higher-dimensional replacement for the one-dimensional second derivative is the Monge-Amp\`ere operator given by the determinant of the matrix of second derivatives, i.e. the Hessian. Denoting this operator by $\operatorname{MA},$ the result of applying it to a tropical series $F$ is a sum of Dirac delta's supported in the vertices of $V(F),$ i.e. $$\operatorname{MA}(F)=(-1)^d\sum m_{v,F} \delta_v,$$ where the multiplicity $m_{v,F}$ is the volume of a polyhedron spanned by the gradients of tropical monomials minimizing $F$ at $v.$ Specifically for $d=2,$ we know that $V(G_{p_1,\dots,p_N}0_\Omega)$ is a non-singular (this terminology and broader context are briefly explained, for instance, in \cite{brugalle2015brief}) tropical curve for generic $p_1,\dots,p_N$ and $\Omega$ having smooth boundary, see \cite{kalinin2018introduction}, i.e. all the vertex multiplicities are equal to $\frac{1}{2}.$ Relaxing the smoothness of the boundary, the vertices with higher multiplicity are still sparse. Nevertheless, for non-polygonal $\Omega$ the total number of vertices is infinite. However, this number is finite in every compact region contained in the interior of $\Omega.$

Now, we consider the graph $\Gamma_N$ given by removing all the infinite branches of $V(G_{p_1,\dots,p_N}0_\Omega),$ i.e. leaving only those edges which are contained in a minimal non-trivial cycle. One of the observations of \cite{kalinin2023some} is that $\Gamma_N$ has exactly $N$ cycles since removing the points $p_1,\dots,p_N$ from it gives a tree. In addition, the general theory of tropical optics implies that $\Gamma_N$ is trivalent \cite{mikhalkin2024wave,shkolnikov2025planar}. Therefore, denoting by $\operatorname{vert}_N$ and $\operatorname{edge}_N$ the numbers of vertices and edges of $\Gamma_N$ respectively, we have $3\operatorname{vert}_N=2\operatorname{edge}_N$ due to trivalency, and $\operatorname{vert}_N-\operatorname{edge}_N=1-N,$ which together give $\operatorname{vert}_N=2N+\operatorname{branch}_N-2,$ where $\operatorname{branch}_N$ is the number of terminal branches that has been removed, it has order $O(\sqrt{N})$ and so it vanishes in the scaling limit.

One may expect that the density of vertices of $\Gamma_N$ is proportional to that of the perturbation points. Indeed, this is supported both by numerical simulations, whose results will be documented in a forthcoming work, and by theoretical derivation of \cite{KLSS}, and one has $$\operatorname{MA}(G_{p_1,\dots,p_N}0_\Omega)\approx \delta_{p_1}+...+\delta_{p_N}.$$ The right-hand side due to Monte-Carlo integration principle converges to $\rho$ after multiplying by $N^{-1}.$ However, in dimension $d=2,$ the operator $\operatorname{MA}$ scales quadratically, i.e. $\operatorname{MA}(rf)=r^2\operatorname{MA}(f).$ Therefore, we need to consider $N^{-\frac{1}{2}}G_{p_1,\dots,p_N}0_\Omega$ as an appropriate normalization, which by the above converges to the unique concave solution of the equation $\operatorname{MA} f=\rho$ with the Dirichlet boundary condition on $\Omega.$ We denote this solution by $F_{\rho,\Omega}.$

Since the Monge-Amp\`ere operator is equi-affine invariant, we have $$F_{A(\Omega),A_*(\rho)}=F_{\Omega,\rho}\circ A^{-1},$$ for any $A$ in $\operatorname{SL}_2(\mathbb{R}).$ Thus, the symmetry emergence is established. Finally, we would like to make a remark that the choice of the initial series $0_\Omega$ for performing the infinite perturbation is irrelevant, the rescaled tropical toppling function converges to the same critical state $F_{\Omega,\rho}.$

For a rigorous and detailed proof of this second Tropical-to-Affine scaling limit, see \cite{KLSS}.

\section{From macro to micro scale}

Now we would like to illustrate how solving the system macroscopically allows to draw consequences at the microscopic level. To do so, we start with the most concrete example of $\Omega$ being the unit disc given by $x^2+y^2\leq 1$ and the perturbation profile being uniform, i.e. $\rho(x,y)=\pi^{-1}.$ The solution of the Monge-Amp\`ere equation thus is $F=\frac{1}{2\sqrt{\pi}}(1-x^2-y^2).$ At the Abelian sandpile level this implies that, provided $h$ is small and $N$ is large, the toppling function of the uniform perturbation of the state $\langle 3 \rangle$ at $N$ uniformly chosen points on $h$-discretization of the disc is expected to be close to $h^{-1}\sqrt{N}F.$ To estimate the deviation of density of the perturbed state on a square window $S$ inside $\Omega,$ we compute the integral over $S$ of the Laplacian of $h^{-1}\sqrt{N}F$ multiplied by $h^2$ and divided by the area of $S,$ which gives $\frac{h\sqrt{N}}{2\pi}$ (by affine symmetry same analysis holds for all ellipses). In other words, we may expect that the perturbed state on the disc is uniformly dense provided that $N<h^{-2},$ giving in addition the threshold for the validity of our regime, since the density must be a positive quantity. It should be noted that this example is not within the reach of the theory as the density's support is not disjoint from the boundary. However, numerical simulations agree with the above derivation.

\begin{figure*}
    \centering
    \includegraphics[width=\linewidth]{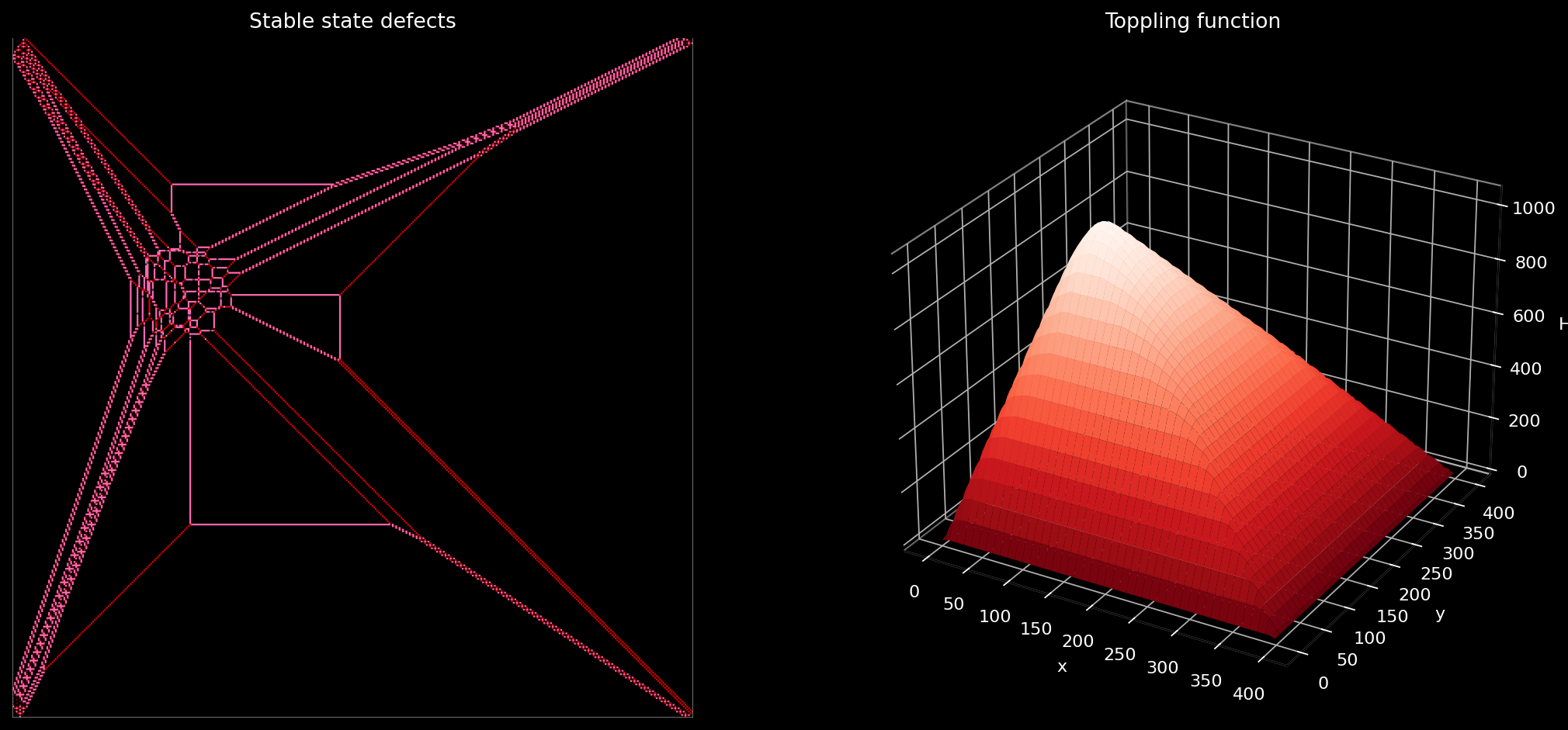}
    \caption{A perturbation by a localized cloud. Observe that the plot on the right is near-conical over the center of the perturbation, and the corresponding set of defects on the left has almost no vertices outside the perturbation cloud. }
    \label{fig:delta_cloud}
\end{figure*}

In the opposite extreme of the uniform perturbation, is the perturbation confined to a mesoscopic vicinity of a point $p$ in the interior of  $\Omega,$ which now is assumed to be a general compact convex domain. Macroscopically, this given by the single-point distribution $\delta_p,$ which at microscopic level should be implemented, nevertheless, by a cloud of generic perturbations near $p.$ On the one hand, this is the setup of applicability of our scaling limit theorems, and, indeed, it is easy to provide anomalous perturbations: take for instance a sequence of perturbation points approaching $p$ horizontally -- the resulting toppling function will be very different from the predicted solution of the Monge-Amp\`ere equation, whose plot is a cone over $\Omega$ with $p$ being the projection of the apex. Repeating the analysis of the previous paragraph, we see for a strictly convex domain $\Omega$ the perturbed state has a non-trivial deviation everywhere, and if the domain has a straight segment on the boundary $s,$ the density deviation vanishes on the sector of $\Omega$ spanned by $s$ and $p,$ see Fig \ref{fig:delta_cloud}.

At a more mathematically precise level, we put forward the following conjecture. Fix $c>0$ and $\alpha>\frac{1}{2},$ assume that $h_N$ is now expressed in terms of $N$ as $cN^{-\alpha}.$ For every natural $N,$ consider states $\phi_N$ on $\Omega_{h_N}$ given by $(\langle 3 \rangle+\sum_{p\in P_N}\tilde\delta_p)^\circ,$ where $P_N$ is an $N$ element subset of $\Omega_{h_N}$ such that $N^{-1}\sum_{p\in P_N}\delta_p$ converges to a given probability density $\rho$ whose support is contained in $\Omega$ and disjoint from the straight segments of its boundary (if there are any). Then, there exists a weak-* limit $E_\rho$ of the rescaled deviation $N^{\frac{1}{2}+\alpha}(\langle 3 \rangle-\phi_N)$ as $N\rightarrow\infty.$ Moreover, $E_\rho$ is characterized by the property that the solution $g$ of the Poisson equations $\Delta g=-c^{-1}E_\rho$ serves as the unique concave solution to the Monge-Amp\`ere equation $\operatorname{MA} g=\rho,$ where in both cases one assumes Dirichlet boundary conditions.

\section{Origins and development}
As it is often the case, it is hard to trace back all the inputs the above story has had. Nevertheless, we feel the obligation and appropriateness to honor the key instances and influences that made this note possible.

The first episode in the local history of these ideas arose from a highly improbable coincidence: sometime around 2014, Grigory Mikhalkin and Andrea Sportiello happened to be lecturing in Paris at the same time. Both have recognized that they used similar kind of pictures to communicate their ideas. To the first these were called tropical curve, and for the second one these were sandpile strings. 

Despite different apparent natures and formalisms, it became clear very soon that there is a mechanism for transforming one to another, namely the Abelian-to-Tropical scaling limit theorem, inspired by the at-that-time recently established scaling limit \cite{pegden2013convergence}. This observation took its initial public form in February 2015 after less than half a year of intense work by the second author and Nikita Kalinin, both of whom were PhD students of Grigory Mikhalkin at that time. This text went through numerous transformations and splittings, and only after 11 years was it published \cite{kalinin2026tropical}, mostly due to the inadequacy of writing experience that the two authors shared.

Among the very first to recognize the potential of that work, after being exposed to it at a talk of the second author in Oberwolfach in 2015, was the first author, who generously shared his advice and gave encouragement for this stream of thought to continue. One of the earlier fruits of this collaboration was \cite{kalinin2018self}, where the notion of tropical sandpile has emerged. In parallel, there two works by Nikita Kalinin and the second author were published, where first steps in the number-theoretic implications of the tropical series framework were made -- see \cite{kalinin2017number, kalinin2019tropical}. An important observation was made by Fedor Petrov, who showed that Dirichlet generating series for critical times in the tropical wave front of a disk is convergent for $\Re{s}>\frac{2}{3},$ and diverges at $s=\frac{2}{3}.$

Understanding the geometric meaning of this singularity became a central question in Nikita Kalinin’s work and was ultimately resolved in the joint work \cite{Residues}, where the residue is shown to be universally proportional to the equi-affine perimeter. Importantly, the corresponding residue is universally proportional to the equi-affine perimeter of the $C^3$ boundary convex domain -- this was the first instance of a bridge between tropical and affine shores that we have personally encountered. After attending one of his earlier talks in the beginning of 2025 in Hong Kong, Conan Leung has made several comments on a potential deeper underlying correspondence between the two geometries. The first interpretation of what he suggested was a trick common in theoretical physics consisting of averaging out the choices in order to achieve more symmetries (think for, instance, about averaging quantities in a QFT over choices of Riemannian metrics to arrive to a TQFT); in this particular case such a choice is underlying tropical structure -- this idea is still not fully developed, but some basic results and descriptions of open problems in this direction may be found in a recent note \cite{kalinin2026limits}.

In an indirect way, this provided a first vision of the setup for the present note, when the two authors were at a hike on Vitosha Mountain in August 2025 contemplating an idea of an infinite perturbation of a tropical sandpile -- here the earlier work of Nikita and Yulieth Prieto was an important source of inspiration and guidance. After the first experiments were made, it was clear that the averaged distances and the uniformly perturbed states share a lot of features, the most striking of which would be the affine covariance. For the fall semester of 2025 most of the experiments were made by Alex Varypaev, who confirmed statistically the convergence and concentration of measure for the growing numbers of perturbation points distributed according to a fixed density.

Nevertheless, the conjectural mechanism for the affine symmetry emergence was still missing. The first author in parallel, was suggesting that there should be a differential equation describing the infinetely perturbed state, which was provoked by looking at the trivial one-dimensional case, as well as by higher-level D-module considerations.

It wasn't clear what this equation is until the second author read the updated introduction of the above-mentioned averaging paper \cite{kalinin2026limits} with Nikita Kalinin before its submission, where some references to Monge-Amp\`ere equation were given. After that, Alex Varypaev made a few more experiments, evaluating the MA operator on infinitely perturbed states, that supported the discovery described in the main part of this note. The theoretical mechanism presented here by now was thoroughly tested by Higinio Serrano and has very recently taken the form of a mathematical theorem \cite{KLSS}.

\begin{acknowledgments}\vspace{-8pt}
This work was supported by the Simons Foundation under grant
SFI-MPS-T-Institutes-00007697 and by the Ministry of Education and
Science of the Republic of Bulgaria under grant DO1-239/10.12.2024.
E.L. gratefully acknowledges Cinvestav for a sabbatical leave during
which most of this work was carried out.

We are grateful to the growing community surrounding these ideas,
especially to the students whose enthusiasm, diligence, and willingness
to explore across disciplinary boundaries continually open new
directions. We also thank the colleagues, institutions, and supporters
who share and sustain the broader aim of fostering meaningful exchanges
among mathematics, physics, and the other forms of human creativity.
\end{acknowledgments}

\bibliography{infinetelyperturbed}

{\centering
\vspace{8pt} 
The Center for Research and Advanced Studies\\ of the National Polytechnic Institute (Cinvestav)\\
Av. Instituto Politécnico Nacional 2508\\
Col. San Pedro Zacatenco, Alcaldía Gustavo A. Madero\\
Mexico City 07360, Mexico\\\vspace{4pt}
and\\\vspace{4pt}
Institute of Mathematics and Informatics\\
Bulgarian Academy of Sciences\\
Akad. G. Bonchev, Sofia 1113, Bulgaria \\\vspace{5pt} 

elupercio[at]gmail.com\\\vspace{20pt}

Institute of Mathematics and Informatics\\
Bulgarian Academy of Sciences\\
Akad. G. Bonchev, Sofia 1113, Bulgaria \\\vspace{5pt}

m.shkolnikov[at]math.bas.bg\\

}

\end{document}